# Superconductivity in im-miscible Cu-Nb phase separated nano-composite thin films

[a]Pradnya Parab, [b]Vivas Bagwe, [b]Bhagyashree Chalke, [c]H. Muthurajan, [b]Pratap Raychaudhuri and [a]Sangita Bose

[a]*UM-DAE Center for Excellence in Basic Sciences, University of Mumbai, Kalina Campus, Santacruz (E), Mumbai-400098, INDIA.*

[b]*Tata Institute of Fundamental Research, Department of Condensed Matter Physics and Material Science, Homi Bhabha Road, Colaba, Mumbai-400005, INDIA.*

[c]*National Centre for Nanoscience & Nanotechnology, University of Mumbai, Kalina Campus, Santacruz (E), Mumbai-400098, INDIA.*

Email: sangita@cbs.ac.in

*Abstract: Superconductivity in granular films is controlled by the grain size and the inter-grain coupling. In a two-component granular system formed by a random mixture of a normal metal (N) and a superconductor (S), the superconducting nano-grains may become coupled through S-N weak links, thereby affecting the superconducting properties of the network. We report on the study of superconductivity in immiscible Nb-Cu nanocomposite films with varying compositions. The microstructure of the films revealed the presence of phase separated, closely spaced, nano-grains of Nb and Cu whose sizes changed marginally with composition. The superconducting transition temperature ($T_{c0}$) of the films decreased with increasing concentration of Cu with a concomitant decrease in the upper critical field ($H_{c2}$) and the critical current ($I_c$). Our results indicate the presence of superconducting phase fluctuations in all films with varying Nb:Cu content which not only affected the temperature for the formation of a true phase coherent superconducting condensate in the films but also other superconducting properties.*

**Introduction**

Granular films with arrays of Josephson junctions have been studied since the past two decades but recently there has been a renewed interest in this field [1,2,3,4,5,6,7,8,9]. This is due to some of the important questions pertaining to the destruction of superconductivity of such systems in relation with the coupling between the grains. In single component superconducting granular films in 2D, a superconductor-insulator transition is observed with decrease in film thickness which results in an increase in the sheet resistivity. This has been ascribed to an increase in phase fluctuations between the grains as the coupling between them decreases, while the local pairing of the Cooper pairs remains unchanged [4,10,11]. Recent studies of THz spectroscopy on Al films with 2-3 nm grains show that with a decrease in the coupling between the grains, the superfluid stiffness decreases though the superconducting energy gap ($\Delta$) remains unchanged. These experiments confirms this scenario of increased phase fluctuations as the cause for the destruction of superconductivity in granular Al films. There are very few studies on granular films comprising of superconductor and normal metal junctions probing the role of phase fluctuations on superconductivity. In addition, in these two component granular films, the superconducting proximity effect (SPE) arising due to the diffusion of the electron pairs across the interfaces also become important. Recent experiments performed on 2D graphene decorated with Sn islands have shown that SPE gets suppressed due to quantum phase fluctuations [12]. In addition their experiments also proved the existence of the zero temperature metallic states which forms another interesting aspect in such arrays of Josephson junctions in the 2D limit. Both SPE and the zero temperature metallic states was also probed for the first time in a completely tunable, artificially engineered arrays of Nb islands on gold, 2D films [13]. In addition, the role of phase fluctuations between the Nb islands was also explored. Though both these pioneering studies on these perfectly engineered films capture well the essential features of an array of S-N junctions, they are done for superconducting islands of the order of few hundred of nanometers which are separated by 100-1000 nm. In addition, these studies were done in 2D systems where additional effects like localization can also suppress superconductivity.

Studies on 3D superconducting granular films with nano-interfaces have also been attempted earlier. These systems can be made sufficiently dis-ordered without localizing the electrons or Cooper pairs, thus making them good model systems to study the role of phase fluctuations on the suppression of superconductivity. The earliest studies were done on random mixtures of Al-Ge and Pb-Ge which formed an array of S-I junctions [14,15]. In these systems the superconductor-insulator transition was primarily controlled by the percolation threshold at which an infinite superconducting cluster was formed and there were no experimental signatures for the presence of phase fluctuations. Later there were some reports on random arrays of S-N junctions [9]. In these systems, disorder could be tuned by changing the relative content of the two (N and S) but the normal state always remained a good metal. Studies on Pb-Cu, Pb-Ag granular films showed that the $T_c$ varied with the relative content of the two constituents [16,17,18]. In these films, the grain sizes were lower than the respective coherence lengths, and it was observed that the $T_c$ followed a similar variation as predicted (and observed) for bilayers and multilayers [19,20], if the thickness of the two was replaced by the volume fractions of the two. Though excellent agreement was obtained of the experimental data with this model, other parameters like coupling between the grains and the sharpness of the interface was not considered [21].

In this paper, we report on a systematic study of the superconducting properties in the immiscible binary system of Nb-Cu [22,23], grown as 3D granular films which consists of well separated Nb and Cu nanoparticles. Transport measurements show two distinct transitions where the lower temperature corresponds to the formation of the phase coherent state via Josephson coupling of the superconducting Nb nanoparticles via the Cu nanoparticles. The higher temperature depends on the size of the Nb grains and is primarily influenced by quantum size effects (QSE) observed in small superconductors [24,25]. We attribute the 2$^{nd}$ transition to the establishment of long-range phase coherence between grains. Both the temperatures decreases rather slowly with increasing Cu content which cannot be accounted for by SPE alone. The role of superconducting phase fluctuations is further confirmed through measurements of the critical fields and critical current which also decrease with increasing Cu content. Our study shows that in such 3D S-N nano-composite films, superconductivity is influenced by the presence of phase fluctuations between the superconducting grains.

**Experimental Details**

The granular thin films have been grown by DC magnetron co-sputtering of Nb and Cu. The sputter deposition was carried out in a custom-built chamber in which the Nb and Cu targets were mounted in a con-focal geometry facing the substrate. The targets used for the sputter deposition were commercially purchased elemental Nb and Cu of 99.99% and 99.95% purity respectively. All depositions were carried out on oxidized Si [100] (p-type) substrates which had a thick (~200 nm) amorphous $SiO_2$ on the surface. The pressure of the Ar gas during sputtering was mostly kept at 5.3 x$10^{-3}$ mtorr. Some depositions were also done at slightly higher pressure of 1.0 x $10^{-2}$ to 25 x $10^{-2}$ mtorr to tune the particle size of Nb. The DC power for Nb deposition using a 2″ diameter target was kept at ~180-202 W for most of the films. The substrate temperature was kept at 600 $^0$C. In these films, the ratio of the Nb:Cu was varied by changing the ratio of the power densities (Sputtering power/ area of the target) of the two and they were grown for 5 mins which resulted in 3 dimensional films of thickness 150-250 nm. In this series, some films were also grown with lower sputtering power of Nb (~120 W).

All the as-deposited Nb-Cu granular films or nano-composites have been characterized using x-ray diffraction (XRD) in a 1.6 kW Rigaku machine using CuK$\alpha$ ($\lambda$ = 1.54 Å) beam. The surface morphology and the elemental composition were estimated using scanning electron microscopy (SEM) and energy dispersive x-ray analysis (EDAX) using a field emission scanning electron microscope (FESEM) operated at 20 kV (Zeiss Ultra FESEM, Germany). The particle size (*d*) was determined in terms of the coherently diffracting domain size from XRD line broadening using the Debye-Scherrer formula after correcting for instrumental broadening. The particle size and the inter-granular region was further characterized using a FEI TITAN Transmission electron microscope (TEM) operating at 300 keV. EDAX analysis was also done at a very local scale using the high annular dark field (HADF) imaging mode of the machine. The superconducting transition temperature ($T_c$) of the films was measured using a two-coil mutual inductance technique [26] integrated with a sample in vacuum 2.8 K cryogen free system (Cryo Industries of America). Transport measurements were also carried on the films in the standard four probe geometry to measure the temperature variation of resistivity as well as the temperature dependence of the critical currents ($I_c$). The critical fields ($H_{c2}$) of the films was estimated from the magnetic field variation of the mutual inductance at different temperatures. Point contact Andreev reflection (PCAR) spectroscopy was done to measure the superconducting energy gap ($\Delta$) of the films. It was done in the conventional needle-anvil technique in which a ballistic contact was made with a Pt-Ir tip on the superconducting film [27]. Current-Voltage (*I-V*) characteristics of the junction was measured using a Keithley 2400 universal sourcemeter at different temperatures. These were numerically differentiated to obtain the PCAR spectra (conductance (G(V) = dI/dV) *vs* the voltage (V)). These spectra are fitted with the Blonder-Klapwijk-Tinkham (BTK) theory [28] used to model normal metal-superconductor interfaces to obtain Δ.

**Results**

Figure 1(a)-(b) show representative XRD spectra acquired for the films grown at $600^0$C for different ratios of the sputtering power densities of Nb:Cu. All films show distinct peaks corresponding to BCC Nb [110] and FCC Cu [111] indicating phase separation of the two immiscible elements. The Cu content in the films could be tuned monotonically from 0% to 75% which was estimated from an EDAX analysis. The grain size of Nb ($d_{Nb}$) estimated from the line broadening of XRD using Debye-Scherrer formula (also called the x-ray domain size), changes only slightly with increasing Cu content in the films and ranges between 16-10 nm. The x-ray domain size for Cu ranges between 8-20 nm. For some films, the particle size was tuned by increasing the Ar pressure during deposition keeping a fixed ratio of the sputtering power of Nb:Cu. For such films, XRD showed considerable broadening of the Nb[110] peak (Figure 1(b)) indicating further coarsening of the grain size. Films with $d_{Nb}$ < 6 nm were also non-superconducting. However, all films had a crystalline, granular structure as seen from the microstructure of the SEM images (SEM of two representative films is shown in Figures 2(a)-(b)). However, one cannot distinguish between the Nb and Cu grains from these images. A compositional image of the nanoscale two-phase dispersion was successfully obtained using an energy selective backscatter (ESB) detector. Figure 2(c) shows a SEM image along with its ESB image for one of the nano-composite film, where the Nb and Cu particles display different contrasts. The particle size was also confirmed through TEM measurements. A representative high annular dark field (HADF) TEM image on the film with Nb 85At% is shown in Figure 2(d). Small grains of 10-15 nm is seen in the image. A local scale EDAX analysis shows the darker grains to be Nb rich confirming the nano-scale phase separation in the films. A further high resolution TEM image (HRTEM) (Figure 2(e)) reveals small grains (~11nm) separated by ~1 nm inter-granular di-ordered region indicating the absence of sharp interfaces between the grains. It is further seen that pure Nb films grown under the same conditions as used for the nano-composite films has the same grain size ( ~ 15 nm ± 2 nm).

Figure 3(a) shows the plot of the temperature dependence of the resistivity for the Nb-Cu films with varying Nb content. The resistivity gradually decreases with decreasing Nb content and gradually approaches the resistivity of bulk Cu films. Besides all films showed a positive temperature coefficient indicating the formation of metallic films with good connectivity between grains. The presence of the metallic Cu matrix makes the Nb-Cu system distinctly different from the well-studied granular Al-$Al_2O_3$ system which shows a metal to insulator transition accompanied by an increase in the room temperature resistivity with decreasing amount of the superconducting material. Another distinction between the normal metal-superconductor nano-composite films studied here from the conventionally studied superconductor-insulator granular films is brought out through the transport measurements at low temperatures. From the R-T plots at low temperature (Figure 3(b)), close to the superconducting transition, it is observed that the resistance of the films drops to zero in two steps. Interestingly, the two transitions are absent in the mutual inductance measurement (M-T plot seen in Figure 3(c)) which measures the diamagnetic shielding response of the entire film. Three noteworthy observations can be made from Figures 3(b)-(e).

(1) The temperature at which the resistance goes to zero in R-T (called $T_{c0}$) corresponds to the onset of the drop of the diamagnetic shielding response in M-T (see Figure 3(b)-(c)).

(2) The two transitions are most clearly visible in the film with highest Nb content (figure 3(d)). The reason for this could be the low resistivity of the films with increasing Cu content which makes it difficult to distinguish clearly the small drop in resistance. However, taking a derivative of the plots clearly shows the two transitions in the films (Figure 3(e)). The first peak in the derivative plot occurs at $T_c^{ON}$ which is the onset of the superconducting transition in R-T. The second peak occurs at a temperature *T1* and occurs when the resistance drops by only 90 -70% of the normal state value ($R_N$) for all the films.

(3) The two temperatures, $T_c^{ON}$ and $T_{c0}$ decreases with decreasing Nb content.

It is worthwhile to note that the two transitions observed on R-T is similar to that observed in Ref [13] which consisted of ordered arrays of S-N junctions in 2D films and in Ref [9] which consisted of S-N random arrays in 3D films. In these work the lower temperature was associated to the temperature where the superconducting phase was locked across the array formed by Josephson coupled superconducting islands/grains while the higher temperature was associated with the temperature where individual islands/grains became superconducting. In order to ascertain if our system consisting of random mixtures of nano-grains of Nb and Cu forms a random array of Josephson weak links, we carried out *I-V* measurements of all the films. We restricted the measurements at temperatures where no visible heating effect was present. Figure 4(a) shows the temperature dependence of the critical current, $I_c$ (extracted from the *I-V* measurements) for a few representative films. We observe that $I_c$ decreases with increasing Cu content in the films. In Figure 4(b), we plot the normalized critical current *($I_c/I_{c0}$)* with respect to the reduced temperature *($T/T_{c0}$)* for all films shown in Figure 4(a). All plots collapse to a single curve. According to the Ambegaokar-Baratoff theory for a dis-ordered array of well-coupled Josephson junctions [29, 30], the simple relation of $I_c = I_{c0} ((1 - (T/T_c)^4)$ can explain the temperature dependence of the critical current. We tried to fit our data with the above relation and obtained a reasonable fit (solid black line in Figure 4(b)) indicating the formation of an array of Josephson weak links. We also measured the critical fields ($H_{c2}$) of the films by measuring the mutual inductance (M) as a function of the magnetic field (H) . The phase diagram ($H_c$ vs $T$) for some of the films is shown in figure 4(c) which shows the gradual decrease in the critical fields with increasing concentration of the normal metal.

The evolution of the superconducting energy gap ($\Delta$) with composition was also studied by point contact Andreev reflection (PCAR) spectroscopy. Figure 5(a) shows a representative dI/dV-V spectra for the film with Nb 76At% at different temperatures below $T_{c0}$. The value of $\Delta$ was determined by fitting the spectra with the Blonder-Tinkham-Klapwijk (BTK) theory[21] with $\Delta$, Z (barrier parameter), and $\Gamma$ (broadening parameter [22]) as fitting parameters [23]. Reasonable fits were obtained at all temperatures and $\Delta$ showed BCS variation with temperature (see the Figure 5(b)). Figure 5(c)-(d) show the PCAR spectra acquired at low temperatures ( $T/T_{c0} \sim 0.35$-$0.4$) for two nano-composite films with Nb 87At% and Nb 65At% respectively along with the fits from the BTK theory. The PCAR measurements shows that the nano-composite films remain a BCS superconductor with increasing amounts of Cu with the value of $2\Delta(0)/k_B T_{c0} \sim 3.1$-$3.2$.

**Discussions**

The emerging picture from our different measurements is that all superconducting parameters like $T_{c0}$, $T_c^{ON}$, $H_{c2}$ and $I_c$ decrease with increasing Cu content. In addition, $R$-$T$ and $I$-$V$ measurements indicate that our films form a dis-ordered array of Josephson junctions. However, the low resistivity of the films show that they are in a regime where the Josephson coupling energy is much less than the charging energy (Large coupling between the grains). Therefore as opposed to superconductor-insulator nanocomposites quantum phase fluctuations would play a negligible role. Therefore the superconducting properties would be dictated by thermal phase fluctuations and quasiparticle excitations. To further ascertain that the two transitions in our films are associated with the grains and inter-grain coupling respectively, we measured R-T at different magnetic fields for the film with Nb 87At% which showed the two transitions distinctly (See figure 6(a)). In figure 6(b) we plot the temperature variation of the critical fields corresponding to the upper ($T_c^{ON}$) and lower transition ($T_{c0}$). For completeness we also plot the critical fields obtained from the M-H measurements. Interestingly, the H-T phase line corresponding to the lower transition falls below that of the upper line which implies that the superconductivity of the films is killed more rapidly compared to the superconductivity of the individual Nb grains. In Figure 6(b), we also plot the H-T phase line obtained for a pure Nb nanocrystalline film with Nb grains ~ 15 nm which corresponds to grain size of Nb in this Nb-Cu nano-composite film [31]. This confirms that the two transitions observed in R-T measurements in the Nb-Cu nano-composite films correspond to the following: (1) $T_c^{ON}$ corresponds to the temperature when the Nb grains become superconducting. (2) $T_{c0}$ corresponds to the temperature where the overall phase coherence in the films is established. The presence of superconducting phase fluctuations is established in all the films with varying Cu content. However, the coupling between individual Nb grains decreases with increasing Cu content as statistically they become far apart. The decrease in $T_{c0}$, $I_c$ and $H_c$ with increasing concentration of the normal metal (see figure 7(a)) in the films can be attributed to the increase in the destruction of the superconducting phase coherence via the S-N-S Josephson weak links. In Figure 7(b), we plot the variation of $T_c^{ON}$ with Nb content and observe a gradual decrease of only about 2 K as Nb content decreases to 25At% in the films. If $T_c^{ON}$ corresponds to the onset of superconductivity in the Nb grains, it can be affected by the number of Cu grains surrounding it through the superconducting proximity effect (SPE) which should decrease $T_c^{ON}$. However, as the number of S-N junctions in the films increases as the Cu content increases, SPE should increase and decrease $T_c^{ON}$ rapidly, contrary to our observation (Figure 7(b)). It should be noted that many studies have earlier shown that SPE is quenched if the interface between the normal metal and superconductor is rough [21]. The presence of ~ 1-2 nm intergranular region as seen from the HRTEM images in our films indicate the absence of sharp interfaces between the Nb-Cu grains which could decrease SPE. However, we also note that the Nb grain size decreases from ~16 to 8 nm as the Cu content is increased. In Figure 7(c), we plot the variation of $T_c^{ON}$ with $d_{Nb}$ which is very similar to the observed variation of $T_c$ with particle size in nanocrystalline Nb films reported earlier [24,25]. Hence, this decrease in $T_c^{ON}$ in the Nb-Cu nanocomposite films with varying Cu content can be understood on the basis of quantum size effects present in small superconductors (Figure 7(c)). Furthermore, some films with large Nb content (70-80 At%) but with Nb grain size less than 6 nm were non-superconducting which proves that $T_c^{ON}$ is controlled primarily by the Nb nano-grains.

**Conclusions**

In conclusion we report on the study of superconducting properties of 3D random mixtures of S-N nano-composite films. These Nb-Cu films grown by co-sputtering showed distinct phase-separation with formation of nano-grains of Nb and Cu. In all films, with varying content of Nb:Cu, transport measurements reveal that the films approach the zero resistance state through two transitions. The onset of the superconducting transition ($T_c^{ON}$) is controlled by the individual Nb grains. However, when all these individual Nb grains were phase locked via the Josephson coupling through Cu grains, the macroscopic phase coherent ground state was obtained at $T_{c0}$. This was further substantiated from transport measurements in magnetic field, where the transition due to the individual Nb grains ($T_c^{ON}$) evolved much slowly with magnetic field as compared to the phase locking temperature, $T_{c0}$. All superconducting properties like $T_{c0}$, $H_{c2}$ and $I_c$ decrease with increasing Cu content indicating the role of coupling between the grains. Furthermore, the small decrease in $Tc^{ON}$ with increasing Cu content could not be explained solely on the basis of superconducting proximity effect and the variation followed the size variation of $T_c$ arising from quantum size effects in nano-structured superconductors. Our results show the dominating role of phase fluctuations in controlling superconductivity in these random mixtures of S-N nano-composite films.

**Acknowledgements**

The authors will like to acknowledge Prashant Chauhan, Indranil Roy and Sanjeev Kumar for some help in measurements with magnetic field. They would also like to acknowledge R. Bapat, Jayesh and S. C. Purandare for their help in SEM and TEM measurements. We would also acknowledge Mr. Chetan Gurada for providing us access to the XRD facility of the Physics department of the Mumbai University. SB acknowledges partial financial support from the Department of Science and Technology, India through No. SERB/F/1877/2012.

# Figure Captions

**Figure 1**

(a) X-ray diffraction (XRD) spectra of Nb-Cu nanocomposite films with different Nb content with the size of Nb grains > 10 nm. The y-axis is shown in log scale for clarity.

(b) X-ray diffraction (XRD) spectra of representative Nb-Cu nanocomposite films with different Nb content with the size of Nb grains < 6 nm.

**Figure 2:** Representative scanning electron micrograph (SEM) of Nb-Cu films with compositions of (a) Nb 85AtA% and (b) Nb 76At%.

(c) SEM image of the film with composition of Nb 76At%. The left part shows the normal image while the right side shows its backscattered image where Nb grains appear in darker contrast.

(d) HADF image of a film with composition of Nb 85At%. Local EDAX shows grains with darker contrast to be Nb rich and the grains with lighter contrast to be Cu rich.

(e) HRTEM image of the film with composition of Nb 85At% showing grains of ~ 11 nm.

**Figure 3:**

(a) Temperature dependence of resistivity ($\rho$) for the Nb-Cu nano-composite film with varying concentration.

(b) Plot of normalized resistance ($R/R_N$) with temperature for the same set of films as shown in (a) close to the superconducting transition. $R_N$ is the resistance of the films at 10 K. $T_c^{ON}$ is the onset of the transition and $T_{c0}$ is the temperature where the resistance (R) goes to zero.

(c) Plot of susceptibility ($\chi'$) with temperature for the same set of films as shown in (a) close to the superconducting transition.

(d) Plot of R/R$_N$ vs T in an expanded scale for representative Nb-Cu films showing the *two* transitions. The y-axis is plotted in a log scale. For comparison, the R-T plot for a pure Nb film of similar thickness is also shown (a *single* clean transition is observed).

(e) Plot of the derivative of the R-T plot shown in (d) with temperature for the same two representative Nb-Cu films. The scale for the Nb 26At% is shown on the left and that for Nb 85At% is shown in the right. Two peaks showing the presence of two transitions.

**Figure 4:**

(a) Plot of critical current ($I_c$) with temperature for three representative Nb-Cu films. Nb 85At% (blue squares), Nb 65At% (green triangles) and Nb 26At% (orange stars).

(b) Plot of reduced critical current ($i_c/i_{c0}$) with reduced temperature ($T_c/T_{c0}$) for the films shown in (a). The fit with the Ambegaonkar-Baratoff relation (see text) is shown by the black solid line.

(c) Plot of critical field ($H_{C2}$) with temperature for four representative Nb-Cu films. Nb 85At% (blue), Nb 74At% (cyan), Nb 65At% (green) and Nb 26At% (orange).

**Figure 5:**

(a) PCAR spectra (dI/dV vs V) at different temperatures below $T_{c0}$ for a representative Nb-Cu film with Nb76At%. The circles are the data points and the solid lines are the fits using the BTK theory (see text).

(b) Variation of the superconducting energy gap ($\Delta$) obtained from the fits with temperature (solid circles). The black solid line is the BCS variation of temperature dependence of gap.

(c) PCAR spectra for the film with Nb 87At% at T = 2.9 K.

(d) PCAR spectra for the film with Nb 65At% at T ~ 2.4K.

For (c)-(d), the raw data is shown by circles and the solid line shows the BTK fit to the data. The value of $2\Delta/kT_c$ ~ 3 – 3.2 for all the films.

**Figure 6:**

(a) Plot of resistance (R) with temperature at different magnetic fields (shown as legends in the figure) for a representative Nb-Cu film with Nb 87At% showing the evolution of the two transitions with magnetic field.

(b) Plot of critical fields ($H_{C2}$) obtained from (a) with temperature for the Nb 87At% film:

$T_{c0}$ is shown with the bottom and left scale. Here blue circles are from R-T and magenta stars are from M-T.

$T_c^{ON}$ is shown with top and right scale. Here red squares are from R-T. The black half shaded circles are for a Nb nano-crystalline film with $d_{Nb}$ ~ 15 nm (from Ref [31]).

**Figure 7:**

(a) Variation of $T_{c0}$ (magenta stars obtained from M-T and green circles obtained from R-T, the scale is shown in the left) and $H_{c2}$ (red squares, the scale shown in the right) with Nb content (in At%) for the Nb-Cu nanocomposite films.

(b) Variation of $T_c^{ON}$ (obtained from R-T) with Nb content (in At%) for the Nb-Cu nanocomposite films. The red circles are for films with $d_{Nb}$ > 10 nm and the cyan stars are for the films with $d_{Nb}$ < 6 nm.

(c) Variation of $T_c^{ON}$ (obtained from R-T) with $d_{Nb}$. The error bars shows the 15% particle size distribution.

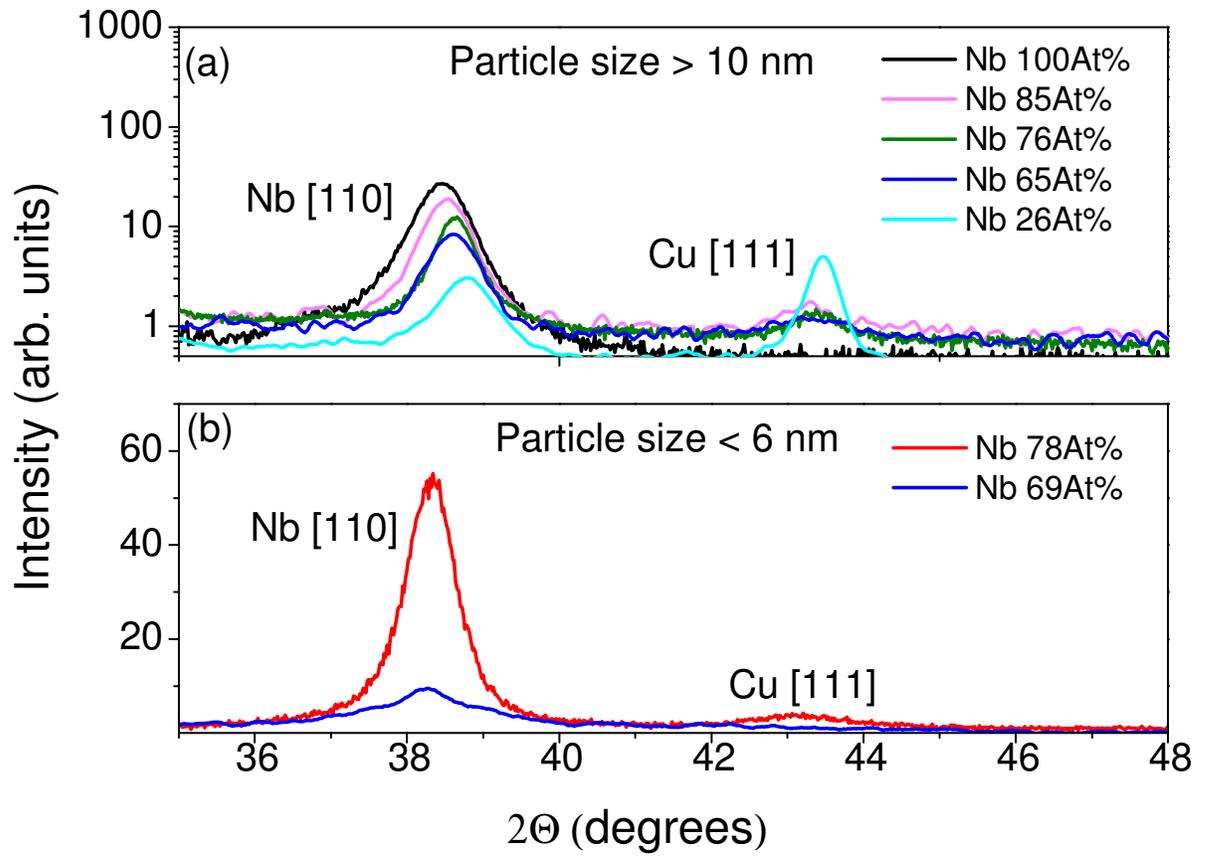

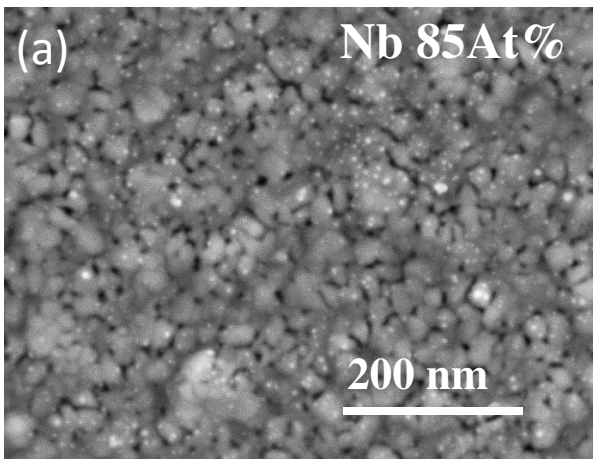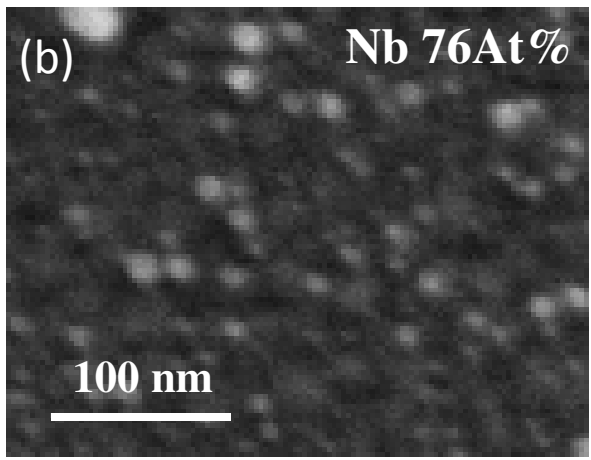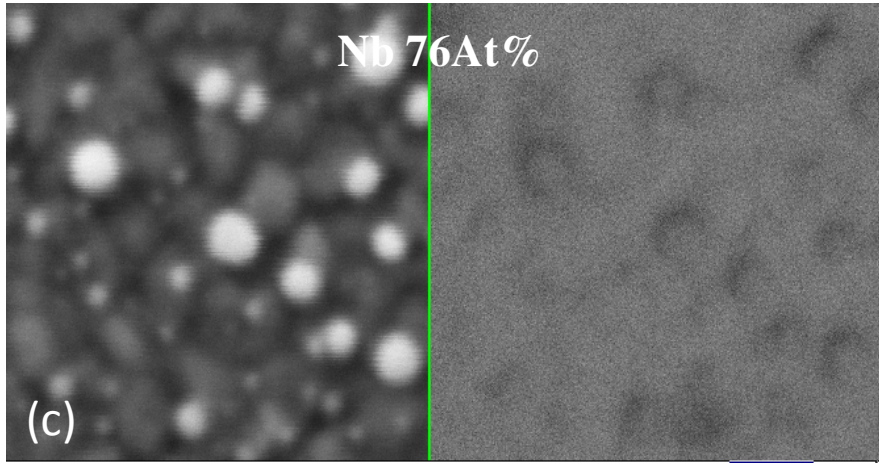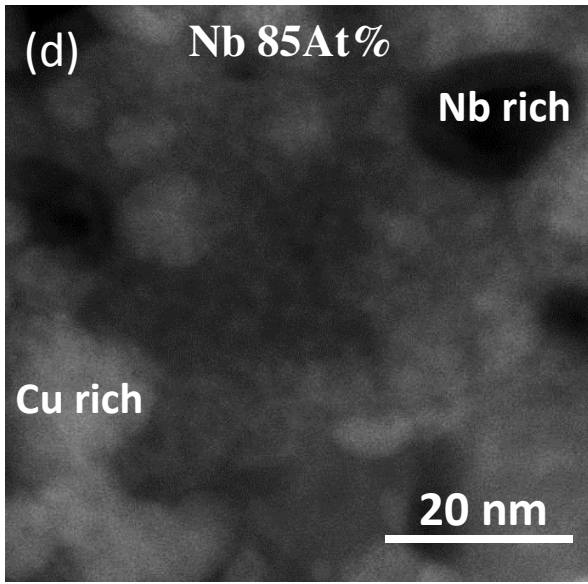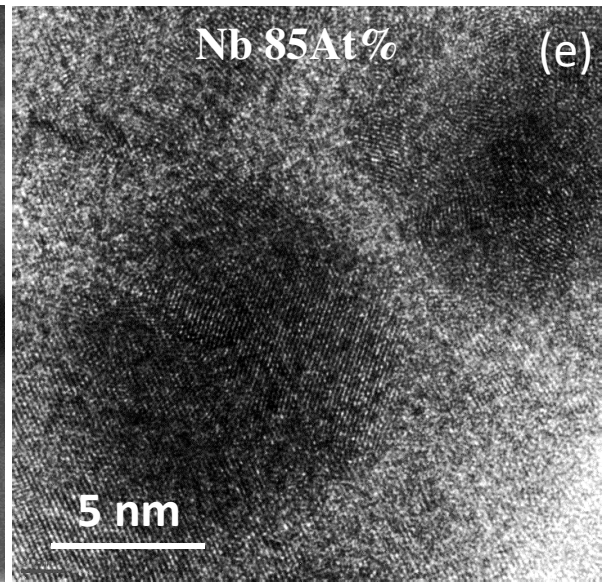

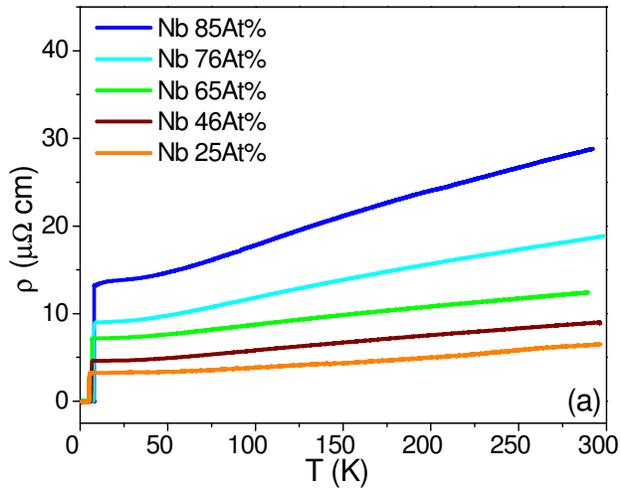
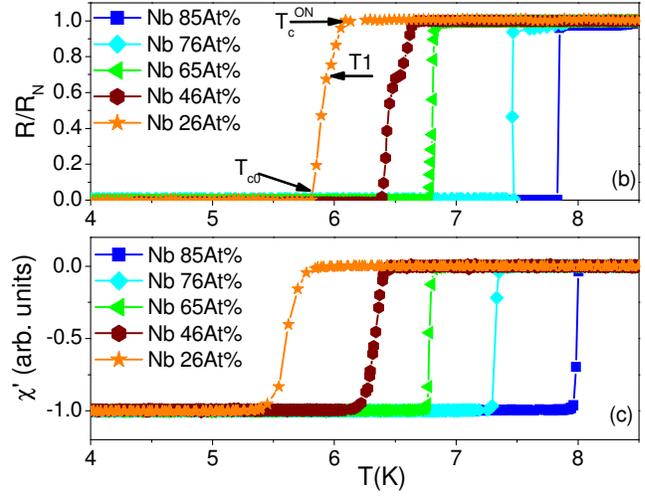
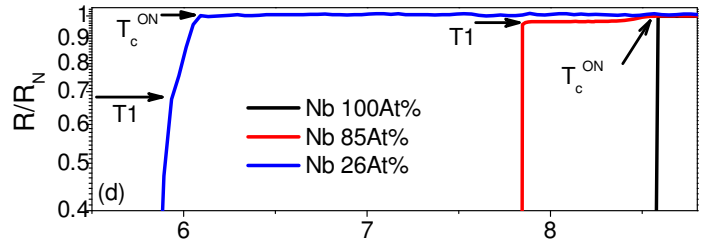
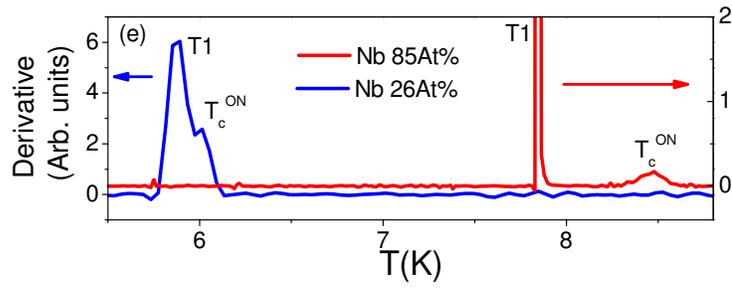

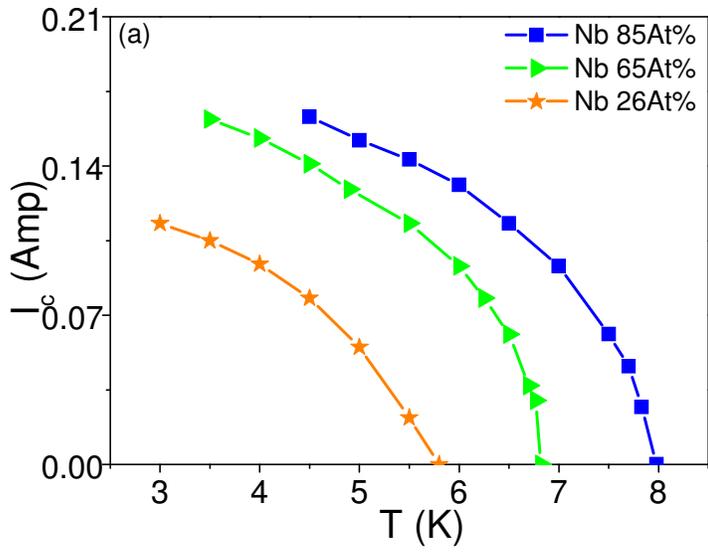
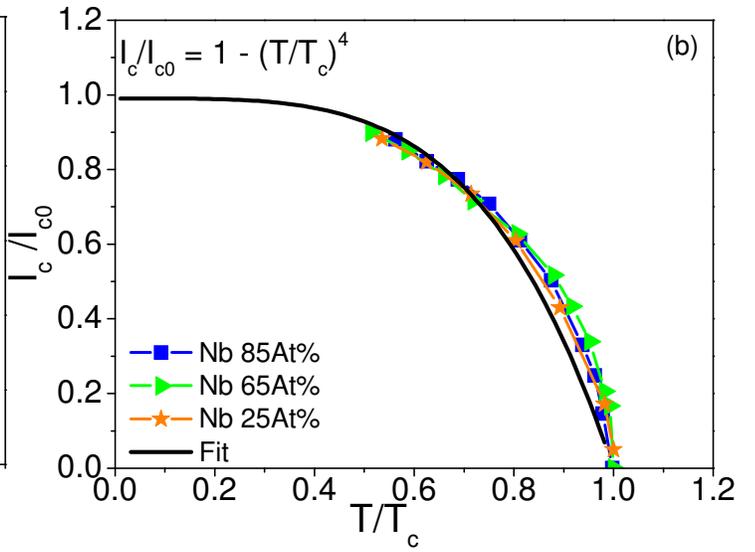
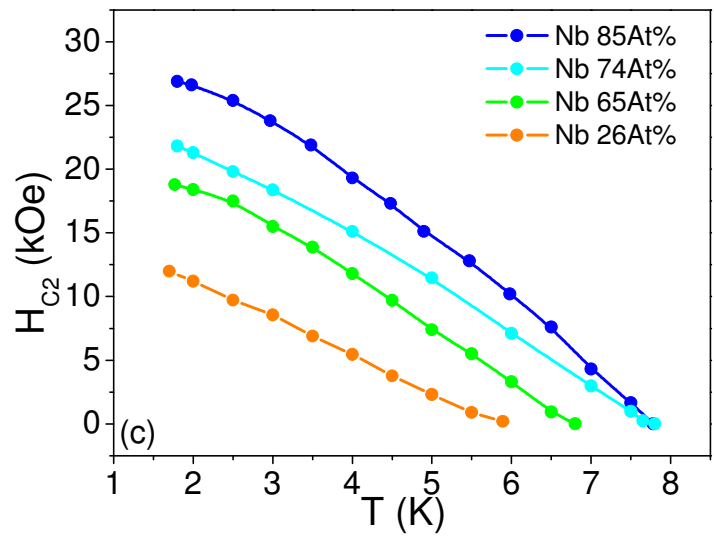

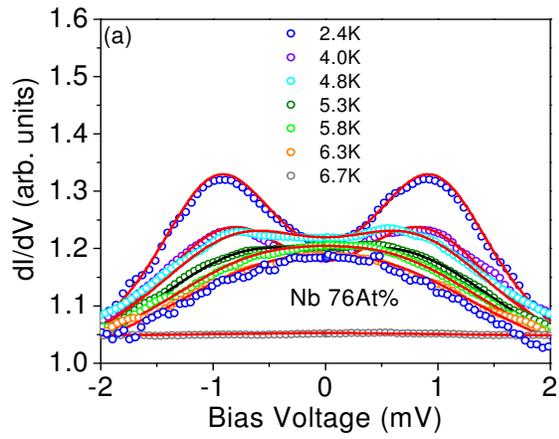
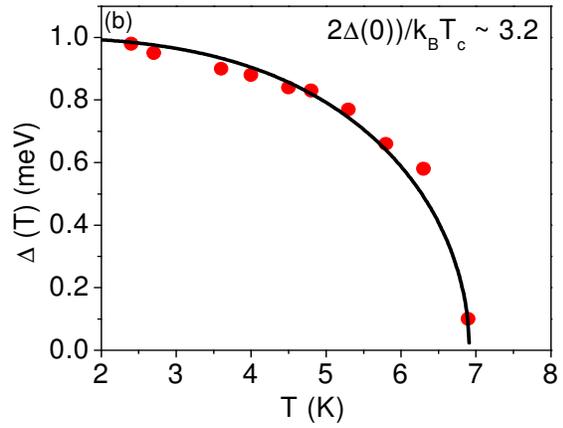
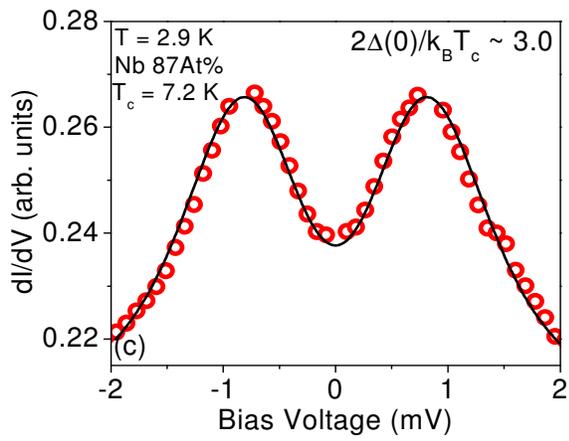
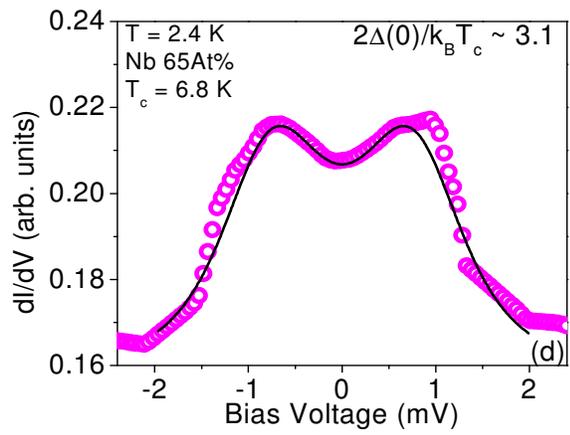

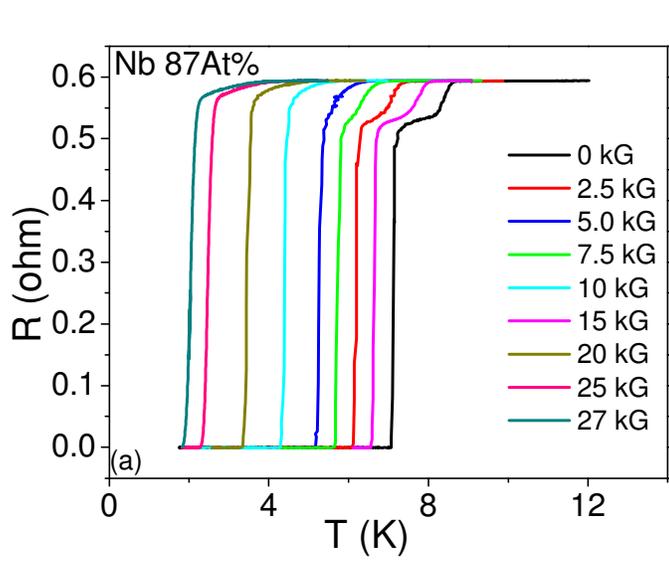 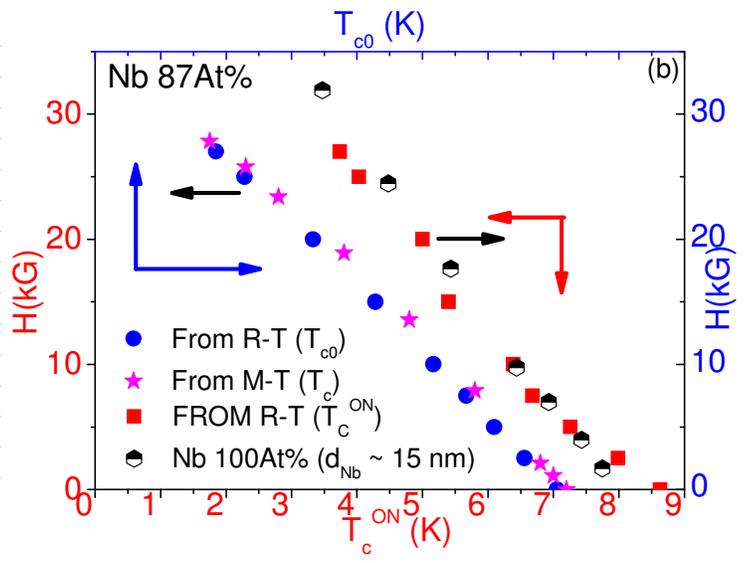

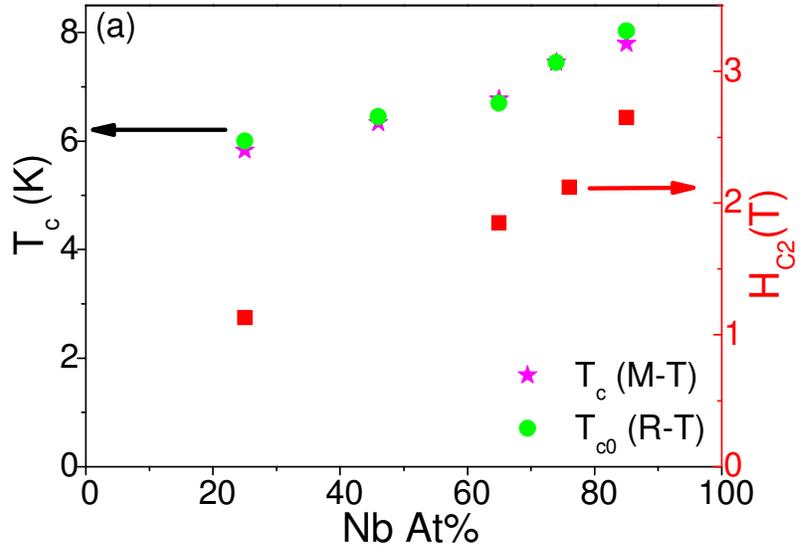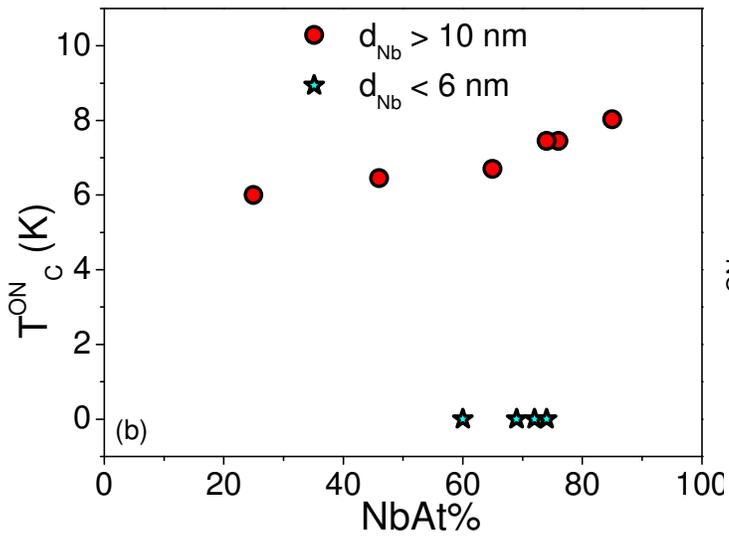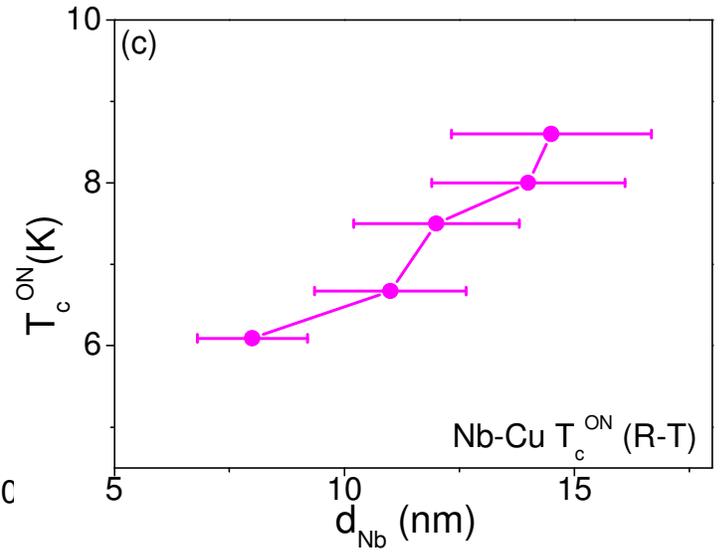